\documentclass[twoside,11pt]{article}
\usepackage{graphics}
\usepackage{psfig}
\usepackage{epsf}
\makeatletter
\newcounter{nbr@exemples}
\newenvironment{exemple}[1]{ \addtocounter{nbr@exemples}{1}
  \def\label##1{\immediate\write\@auxout{\string
    \newlabel{##1}{{\thenbr@exemples}{\thepage}}}}%
    {\immediate\write\@auxout{\string
    \@writefile{exemple}{\string \contentsline\space {figure}{\string
    \numberline\space {\thenbr@exemples}{\string \ignorespaces\space #1}}
        {\thepage}}}}%
    ~\\ \small~\hspace*{0.25in}
    \begin{tabular}{|p{4.25in}}
  \textbf{Example~\thenbr@exemples~(#1)~:} \\
    }{ \end{tabular} ~\\}

\newenvironment{exemple-continu}[1]{
    ~\\ \small~\hspace*{0.25in}
    \begin{tabular}{|p{4.25in}}
  \textbf{Example~#1~(followed)~:} \\
    }{ \end{tabular} ~\\}

\begin{document}
\pagestyle{myheadings} \markboth{WLPE'01}{User-friendly
explanations for constraint programming}
\title{User-friendly explanations for constraint programming
\footnote{In A. Kusalik (ed), Proceedings of the Eleventh Workshop
on Logic Programming Environments (WLPE'01), December 1, 2001,
Paphos, Cyprus. COmputer Research Repository
(http://www.acm.org/corr/), cs.PL/0111037; whole proceedings:
cs.PL/0111042.}}
\author{ Narendra Jussien and Samir Ouis\\
\'Ecole des Mines de Nantes \\
 4 rue Alfred Kastler -- BP 20722 \\
 F-44307 Nantes Cedex 3 \\
email: \texttt{\{jussien,souis\}@emn.fr} \\
web: \texttt{www.e-constraints.net}}
\date{}
\maketitle
\begin{abstract}
In this paper, we introduce a set of tools for providing
user-friendly explanations in an explanation-based constraint
programming system. The idea is to represent the constraints of a
problem as an hierarchy (a tree). Users are then represented as a
set of understandable nodes in that tree (a \emph{cut}). Classical
explanations (sets of system constraints) just need to get projected on
that representation in order to be understandable by any user. We
present here the main interests of this idea.

\end{abstract}

\section{Introduction}
Classical constraint programming systems (such as Solver from
Ilog, Chip from Cosytec or gnuProlog from INRIA) are helpless
when there is no solution to the constraint system to be solved. In fact, only
a \texttt{no solution} message is provided. Users are left alone
to find out why: is it because of the problem itself (no solution
exists), an incorrect modelling, a bug in the solver, etc.

In order to promote constraint programming, the constraints
community needs to address this issue. For example, a set of
constraints that left alone lead to the unexpected situation would
be very informative to the user. Such a set of constraints is
called an \textbf{explanation} \cite{jussien-e-constraints}. It is
a set of constraints justifying propagation events generated by
the solver (value removal, bound update, contradiction). Notice
that even if a lot of debugging tools were developed during the
Discipl European project \cite{deransart-discipl}, there is a lack
for tools which provide explanations.

Explanations (sets of \emph{low-level} constraints) are not
user-friendly: only developers of constraints system understand
them. \emph{Translation} tools are needed. Obviously, input from
the developer of an application is needed. When developing an
application, such an expert needs to \emph{translate} the problem
from the high-level representation (the user's comprehension of
the problem) to the low-level representation (the actual
constraints in the system). We call this translation a
\texttt{user} $\rightarrow$ \texttt{system} translation. For
user-friendly explanations, we need the other way
\emph{translation}: from the low-level constraints (solver
adapted) to the user understandable constraints (higher level of
abstraction). We call this translation a \texttt{system}
$\rightarrow$ \texttt{user} translation. That translation is
usually not explicitly coded in the system. Asking a developer to
provide such a translator while coding would be quite strange. We
chose to automatize that translation in an effortless way.

In this paper, we present an automatic system for generating
user-friendly explanation. We first recall some facts about
explanations within constraint programming. Then, we show how our
system works on hierarchical applications before presenting an
implementation. We conclude this paper with some potential
applications of our user-friendly explanations and some related
works.

\section{Explanations within Constraint Programming}
\label{sec-explication}

We consider here \textsc{csp} represented by a couple $(V,C)$. $V$
is a set of variables and $C$ a set of constraints on those
variables. Notice that variable domains are considered as unary
constraints. Moreover, the enumeration mechanism is handled as a
series of constraints additions and retractions. Those constraints
are called \emph{decision constraints}. Indeed, we chose not to
limit our tools to value assignments but to allow any kind of
decision constraint (\emph{eg.} ordering constraints between tasks
in scheduling, splitting constraints in numeric \textsc{csp}).

Let us consider a constraints system whose current state (\emph{i.e.}
the original constraint and the set of decisions made so far) is
contradictory. A \textbf{contradiction explanation} (\emph{a.k.a.}
\textbf{nogood} \cite{schiex-nogood}) is a subset of the current
constraints system of the problem that, left alone, leads to a
contradiction (no feasible solution contains a nogood). A
contradiction explanation divides into two parts: a subset of the
original set of constraints ($C' \subset C$ in
equation~\ref{eq-nogood}) and a subset of decision constraints
introduced so far in the search (here $dc_1, \ldots, dc_k$).

\begin{equation} \label{eq-nogood}
    \neg \left(C' \wedge dc_1 \wedge ... \wedge dc_k \right)
\end{equation}

An operational viewpoint of contradiction explanations can be made
explicit by rewriting equation~\ref{eq-nogood} the following way:
\begin{equation} \label{eq-explanation}
   C' \wedge \left(\bigwedge_{i \in [1..k] \setminus j}dc_i\right)\rightarrow \neg dc_{j}
\end{equation}

Let us consider $dc_j: v = a$ in the previous formula. The left hand
side of the implication is called an \textbf{eliminating
explanation} (explanation for short) because it justifies the
removal of value $a$ from the domain $d(v)$ of variable $v$. It
will be noted: $\mathtt{expl(} v \neq a \mathtt{)}$.

Filtering operations in \textsc{csp} can be considered as a
sequence of value removals which can all be explained as in
equation~\ref{eq-explanation}. The simplest of all explanations is
to merely consider the complete set of currently active
constraints (\emph{i.e.} the initial constraints of the problem
and the set of all the decisions -- and their associated
enumeration constraint -- made so far). Notice that much more
useful explanations can be provided.

Explanations can be combined with each other to provide new ones.
Let us suppose that $dc_1 \lor \ldots \lor dc_j$ is the set of all
possible choices for a given decision (set of possible values, set
of possible sequences).  If a set of explanations $C'_1
\rightarrow \neg dc_1$, ...,  $C'_j \rightarrow \neg dc_j$ exists,
a new explanation can be derived: $\neg (C'_1 \wedge \ldots \wedge
C'_j)$. Such new explanation gives more information than each of
the old ones.

From the empty domain of a variable $v$, a
\emph{contradiction explanation} can be computed:
\begin{equation} \label{eq-contradiction}
   \neg \left( \bigwedge_{a \in d(v)} \mathtt{expl(} v
    \neq a\mathtt{)} \right)
\end{equation}

Notice that when a contradiction explanation does not contain any
decision constraint, the associated problem is proved to be
over-constrained.

Several eliminating explanations generally exist for the removal
of a given value. Recording all of them leads to an exponential
space complexity.  Another technique relies on \emph{forgetting}
(erasing) explanations that are no longer relevant\footnote{An
explanation is said to be \emph{relevant} if all the decision
constraints in it are still valid in the current search state
\cite{bayardo-complexity}.} to the current variable assignment. By
doing so, the space complexity remains polynomial. We here retain
\textbf{one} explanation at a time for a value removal. Notice
that as explanations reflect the behavior of the solver, a value
in a domain of a variable cannot be removed twice. Therefore, only
one explanation is really computed while solving.

Explanations are useful in many situations
\cite{jussien-palm,jussien-e-constraints}:
\begin{itemize}
\item for debugging problems by providing contradiction
explanation;
\item  for handling dynamic problems by providing the past effects of
constraints;
\item for handling over-constrained problems by combining the two
preceding uses;
\item for defining new conflict-directed search algorithms.
\texttt{mac-dbt} \cite{jussien-macdbt-cp} and \texttt{path-repair}
\cite{jussien-local} are two successful instances.
\end{itemize}

\cite{jussien-e-constraints} introduces the notion of
\emph{e-constraints} to encompass explanations and their use
within constraint programming.

\section{Hypothesis: hierarchical applications}
The work presented in this paper relies on a single hypothesis: all aspects
of a constraint-based application can be represented in an hierarchical way.

\subsection{A problem: an hierarchy of constraints}
Example~\ref{conf-description} presents a small constraint
problem: organizing talks among several people.

\begin{exemple}{The conference problem} \label{conf-description}
Michael, Peter and Alan are organizing a two-day seminar for
writing a report on their work. In order to be efficient, Peter
and Alan need to present their work to Michael and Michael needs
to present his work to Alan and Peter (actually Peter and Alan
work in the same lab). Those presentations are scheduled for a
whole half-day each.

Michael wants to known what Peter and Alan have done before
presenting his own work. Moreover, Michael would prefer not to
come the afternoon of the second day because he has got a very
long ride home. Finally, Michael would really prefer not to
present his work to Peter and Alan at the same time.
\end{exemple}

\begin{figure}[hbtp]
  \begin{center}
  ~\hspace{2.75cm}\epsfbox{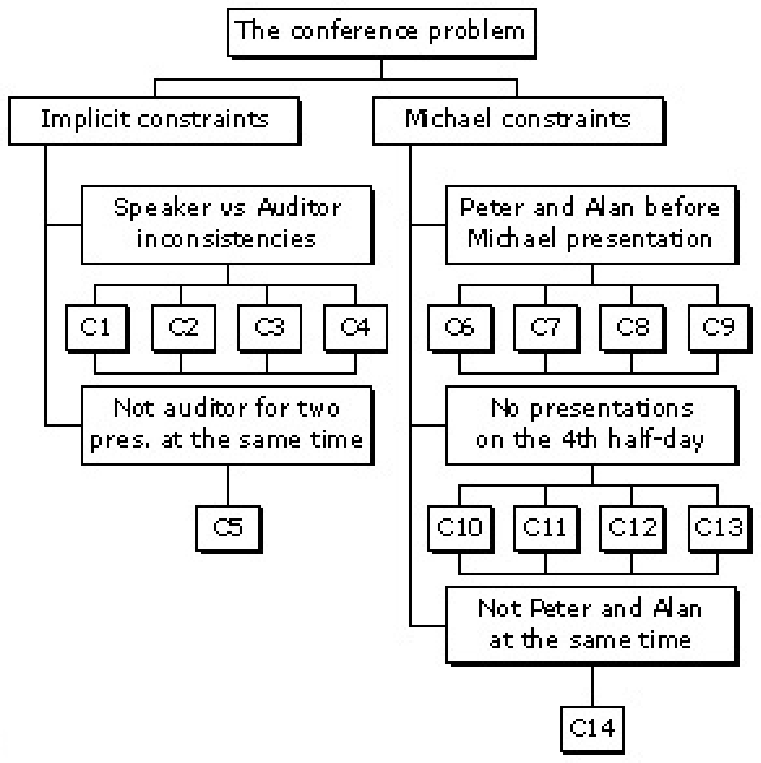}
   ~\vspace{-22.75cm}
  \end{center}
\caption{An hierarchical view of the conference problem} \label{fig-conference}
\end{figure}

A constraint model for that problem is described in
example~\ref{conf-model}. Notice that when modelling the
conference problem, the constraints were categorized leading to an
hierarchy representing the problem. A graphical representation is
presented in figure~\ref{fig-conference}.

Indeed, we claim that it is always possible to attach each
constraint in a given problem to a single father-abstraction. This
general hypothesis may appear as highly restrictive but as we were
trying to find counter-examples we could not exhibit a single one:
we always another way of presenting things that lead to an
hierarchy. We therefore think that our intuition may not be as
restrictive as we thought in the beginning. Moreover, posting
constraint is usually an imperative step in classical constraint
programming. Defining procedures for posting constraints are the
kind of abstraction we are interested in (see
example~\ref{coding-conference}).

\begin{exemple}{A constraint model for the conference problem} \label{conf-model}
Let $Ma, Mp, Am, Pm$ the variables representing the four
presentations ($M$ and $m$ are respectively for Michael as a
speaker and as an auditor and so on). Their domain will be
$[1,2,3,4]$ ($1$ is for the morning of the first day and $4$ for
the afternoon of the second day).

Several constraints are contained in the problem: implicit
constraints regarding the organization of presentations and the
constraints expressed by Michael.

The implicit constraints can be stated:
\begin{itemize}
\item A speaker cannot be an auditor in the same half-day. This
constraint is modelled as: $c_1$: $Ma \neq Am$, $c_2$: $Mp \neq
Pm$, $c_3$: $Ma \neq Pm$ and $c_4$: $Mp \neq Am$.
\item No one can attend two presentations at the same time. This
is modelled as $c_5$: $Am \neq Pm$.
\end{itemize}

Michael constraints can be modelled:
\begin{itemize}
\item Michael wants to speak after Peter and Alan: $c_6$: $Ma > Am$,
$c_7$: $Ma > Pm$, $c_8$: $Mp > Am$ and $c_9$: $Mp > Pm$.
\item Michael does not want to come on the fourth half-day: $c_{10}$: $Ma \neq 4$,
$c_{11}$: $Mp \neq 4$, $c_{12}$: $Am \neq 4$ and $c_{13}$: $Pm
\neq 4$.
\item Michael does not want to present to Peter and Alan at the
same time: $c_{14}$: $Ma \neq Mp$.
\end{itemize}
\end{exemple}

\subsection{Building a \texttt{system} $\rightarrow$ \texttt{user} translator}
While developing a constraint application, the developer only
needs to explicitly state the underlying hierarchy of her problem.
Only the leaves of this structure, namely the low-level
constraints can be used by the constraint solver.

The leaves may be way too low-level for a typical user of the
final application. However, she may understand higher levels in
the hierarchy. The hierarchy hypothesis allows the building, with
no effort for the developer, of an hierarchical representation of
the problem. Once built, this representation may be used to
interact with any user through user-friendly explanations. Such
explanations are provided using procedures converting the
low-level constraints into user understandable nodes of the
hierarchy. Those procedures are completely problem-independent and
may be provided within the constraint solver.

\begin{figure}[hbtp]
\begin{center}
  ~\hspace{2.25cm}\epsfbox{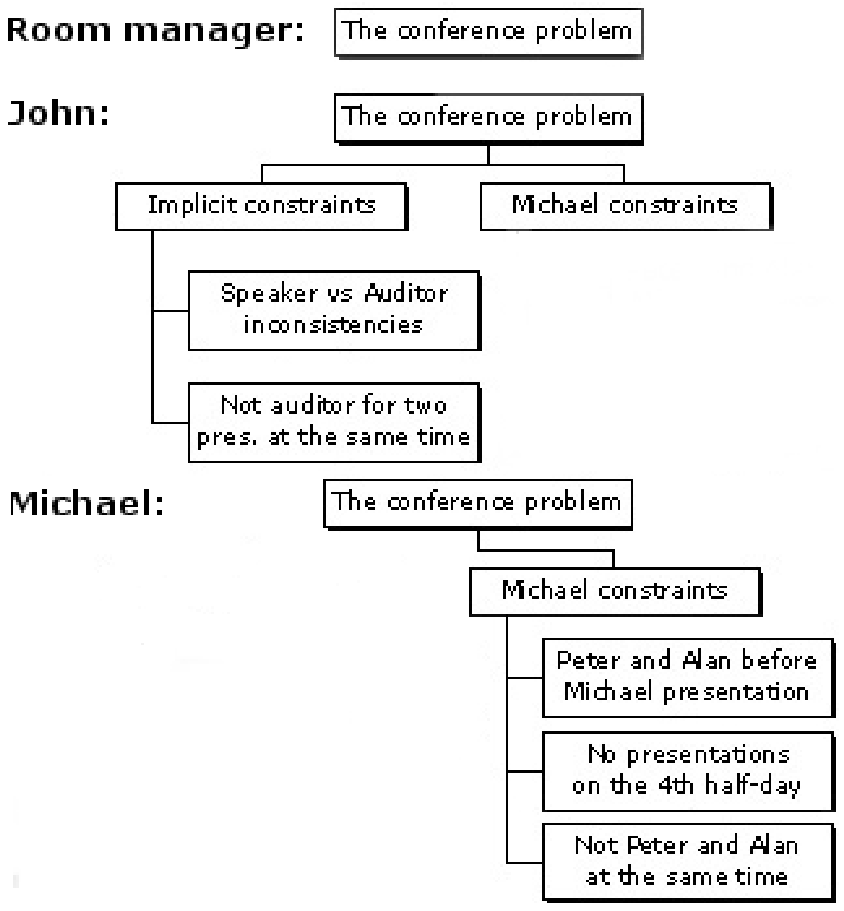}
   ~\vspace{-21.45cm}
  \end{center}
\caption{Different views on the conference problem}
\label{conf-views}
\end{figure}

A user perception of a given problem can be seen as a set of nodes
in that tree (everything above any of those nodes considered as
being understandable and everything below any of those nodes). We
will call that set a {\bf cut} in the hierarchical view of the
considered problem. In our example, here is what it could be (see
also figure~\ref{conf-views}):
\begin{itemize}
\item The room manager of the faculty department has only a very partial
view: she does not want to known about wishes or implicit
constraints. The only part of the problem that she wants to deal
with is the problem as a whole. Therefore, her view of the problem
would be: \fbox{The conference problem}.

\item John who is actually organizing the meetings finds Michael too
complicated. He does not want to deal with his numerous wishes.
But, he does understand the implicit constraints and must deal
with them. Therefore, his view of the problem would be:
\fbox{Speaker vs. Auditor}, \fbox{Auditor vs. 2 pres.} and
\fbox{Michael constraints}.

\item Michael does not want to deal with implicit constraints. Although
he does understand his own wishes. Therefore, his view of the
problem would be: \fbox{The conf. problem}, \fbox{P\&A before},
\fbox{Not 4$^{th}$ $1/2$ day} and \fbox{P\&A not same time}.
\end{itemize}

Computing user-friendly explanations can be done by simply
projecting the low-level constraints in the explanation onto the
user comprehension of the problem in the hierarchy.

Our example, the conference problem has no solution. One
explanation for that situation provided by an e-constraints system
is: $\{c_5, c_6, c_7, c_8, c_9, c_{10},$ $c_{11}, c_{12},
c_{14}\}$.

Here are their translation into user-friendly explanations:
\begin{itemize}
\item For the room manager, the explanation is simple. There is
no possible solution for the problem due to its whole set of
constraints. The projection gives: \fbox{The conf. problem}. He
tells John that there is a problem.

\item John looks at the explanation from his point of view. The projection
gives:

\fbox{Auditor vs. 2 pres.} and \fbox{Michael constraints}. Michael
wishes are too strong because of the \emph{no two presentations at
the same time for a given auditor} constraint. John asks Michael
to review his wishes.

\item Michael looks at the explanation from his point of view. The projection
gives:

\fbox{The conf. problem}, \fbox{P\&A before}, \fbox{Not 4$^{th}$
$1/2$ day} and \\ \fbox{P\&A not same time}. He knows that the
whole set of wishes is a problem. He can choose to discard any.
For example, the constraint on the fourth half-day. This leads to
a solution to the problem.
\end{itemize}

Notice that user input needs also to be translated into low-level
interaction with the constraint solver. A backward projection step
is therefore needed. There are two options: removing all the
concerned constraints or only the constraints that do appear in
the explanation. In our example, we can remove all $c_{10}$ to
$c_{13}$ constraints. But removing $c_{13}$ would be of no use in
our problem, since it does not appear in the explanation.

Moreover, choosing to only remove concerned constraints can help
\emph{partially} enforcing constraints leading to a kind of
\emph{soft constraints} (encoded as a set of possibly removed
low-level constraints).

\section{Implementation: extending the PaLM system}

\subsection{Introducing PaLM}
\texttt{PaLM} is an explanation-based constraint programming
system \cite{jussien-palm} that is provided as a \texttt{choco}
\cite{laburthe-choco} library. \texttt{choco} is the constraint
layer of the \texttt{claire} \cite{caseau-claire} programming
language. \texttt{PaLM} provides tools to handle explanations in a
constraint solver: a specific class, storing methods, retrieving
method, ... \texttt{PaLM} computes explanations while propagating
constraints and can even use them to guide the search
\cite{jussien-e-constraints} (it was used in \texttt{mac-dbt}
\cite{jussien-macdbt-cp} and \texttt{path-repair}
\cite{jussien-local}).

The \texttt{PaLM} system handles variables represented  with a
complete enumerated domain or only by their bounds. It provides
the classical set of basic arithmetic constraints as well as
symbolic constraints (such as \emph{allDifferent}, \emph{element},
...).

\texttt{PaLM} is designed to (automatically) handle
over-constrained problem. If a user wants to define her own
strategy for handling such problems (as one might want to do in
the conference problem), \texttt{PaLM} provides specific
exceptions that can be catched using the standard
\texttt{try/catch} mechanisms of \texttt{claire}.

The \texttt{PaLM} system is is publicly available at
\texttt{www.e-constraints.net}.

\subsection{Tools for user-interaction}
%

\subsubsection{Adding structure information}
The main idea here is to provide tools that allow the less
intrusive possible interaction with the original code of the
application. We therefore introduced the notion of \texttt{UFbox}
(User-Friendly box) that aggregates set of constraints into an
hierarchy. Grouping constraints is done by simply setting the
boundaries of the given boxes using two provided methods:
\texttt{startUFBox} and \texttt{endUFBox}. This explains why we
need the hierarchy hypothesis: code modification is minimal.

Consider the conference problem introduced on
example~\ref{conf-description} and modelled in
example~\ref{conf-model}. Example~\ref{coding-conference} shows an
encoding of that problem in \texttt{choco}.

\begin{figure}[p]
\begin{exemple}{Coding the conference problem with choco} \label{coding-conference}
{\tiny
\begin{verbatim} [conference(): void
 -> let pb := makeProblem("conference problem",4)
        vars := createConferenceVariables(pb)
    in (
       postImplicitConstraints(pb, vars),
       postMichealConstraints(pb, vars),
       solve(pb)
    )]

 // creating the variables
[createConferenceVariables(pb: Problem): list[IntVar] -> ... ]
 // posting Michael constraints
[postMichaelConstraints(pb: Problem, vars: list[IntVar]): void -> ...]

[postImplicitConstraints(pb: Problem, vars: list[IntVar]): void
 -> postSpeakerAuditorIncompatibilityConstraints(pb, vars),
    postNotTwoPresentationsAtTheSameTimeConstraints(pb, vars) ]
// posting the constraint c5
[postNotTwoPresentationsAtTheSameTimeConstraints(pb: Problem, vars: list[IntVar]) -> ... ]

[postSpeakerAuditorIncompatibilityConstraints(pb: Problem, vars:
list[IntVar]): void
 -> post(pb, vars[1] !== vars[2]), // constraint c1
    post(pb, vars[3] !== vars[4]), // constraint c2
    post(pb, vars[1] !== vars[4]), // constraint c3
    post(pb, vars[3] !== vars[2])] // constraint c4
\end{verbatim}}
\end{exemple}
\end{figure}

As you can see, an implicit hierarchy is appearing when encoding
the problem in a programming language: it is easier to maintain
such a program if the constraint posting is structured as it was
during the modelling phase.

In order to use the \texttt{UFboxes} that will be used to
implement the ideas of this paper, one just needs to add some info
while posting constraints. Example~\ref{coding-conf-ufbox} shows
what we get. Notice that \texttt{startUFBox} needs three
parameters: the related \texttt{PalmProblem}, a short description
used to ease user definition (see following section) and a textual
representation of the set of constraints (should be
user-friendly!).

\begin{figure}[p]
\begin{exemple}{Adding UFboxes} \label{coding-conf-ufbox}
{\tiny
\begin{verbatim}
[conference(): void
 -> let pb := makePalmProblem("conference problem",4) // switch to PaLM
        vars := createConferenceVariables(pb)
    in (
       postImplicitConstraints(pb, vars),
       postMichealConstraints(pb, vars),
       setUserRepresentation(pb, list("IC","PAB","N4D","NPA")), // representing Michael
       solve(pb)
    )]
...
[postImplicitConstraints(pb: Problem, vars: list[IntVar]): void
 -> startUFBox(pb,"IC","Implicit constraints"),
       postSpeakerAuditorIncompatibilityConstraints(pb, vars),
       postNotTwoPresentationsAtTheSameTimeConstraints(pb, vars),
    endUFBox(pb)]
...
[postSpeakerAuditorIncompatibilityConstraints(pb: Problem, vars: list[IntVar]): void
 -> startUFBox(pb,"SAIC","Speaker Auditor Incompatibility Constraint"),
       post(pb, vars[1] !== vars[2]), // constraint c1
       post(pb, vars[3] !== vars[4]), // constraint c2
       post(pb, vars[1] !== vars[4]), // constraint c3
       post(pb, vars[3] !== vars[2]), // constraint c4
    endUFBox(pb)]
\end{verbatim} }
\end{exemple}
\end{figure}

\subsubsection{Representing the user}
Tools are also provided to represent the user with the short
descriptions provided while defining the \texttt{UFboxes}: the
\texttt{setUserRepresentation} method that takes a list of short
descriptions to define the cut in the hierarchy tree.

Moreover, projection tools are provided to translate a given
explanation into the current user representation.

Finally, thanks to \texttt{PaLM} capabilities with dynamic
problems, tools are provided for handling dynamic addition or
removal of \texttt{UFboxes} (\emph{i.e.}, sets of constraints as a
single constraint).

\begin{figure}[hptb]
\begin{exemple}{Using UFboxes} \label{ex-use}
{\footnotesize
\begin{verbatim}
palm> conference()
eval[0]> Variables : (Am:[1..4], Pm:[1..4], Ma:[1..4], Mp:[1..4])

=== Conference problem : description
+....[PB] The complete problem
+......[IC] Implicit constraints
+........[SAIC] Speaker-auditor incompatibility constraint
+........[N2P] Not two presentations at the same time
+......[MC] Michael constraints
+........[PAB] Peter and Alan before Michael
+........[N4D] No presentation on the 4th half-day
+........[NPA] Not Peter and Alan at the same time

Solving the problem ...
!!! A contradiction occurred because of :
1: [IC] Implicit constraints
2: [PAB] Peter and Alan before Michael
3: [N4D] No presentation on the 4th half-day
4: [NPA] Not Peter and Alan at the same time

** Which block would you like to relax ? (1-4 0-none)   2
PALM: Removing constraint Mp >= Pm + 1 from PAB
PALM: Removing constraint Mp >= Am + 1 from PAB
PALM: Removing constraint Ma >= Pm + 1 from PAB
PALM: Removing constraint Ma >= Am + 1 from PAB

!!! A contradiction occurred because of :
1: [IC] Implicit constraints
2: [N4D] No presentation on the 4th half-day
3: [NPA] Not Peter and Alan at the same time

** Which block would you like to relax ? (1-3 0-none)   2
PALM: Removing constraint Am !== 4 from N4D

!!! A solution has now been obtained
!!! (Am:4, Pm:1, Ma:2, Mp:3)

!!! The following blocks have been relaxed
  1 : [PAB - 4 cts] Peter and Alan before Michael
  2 : [N4D - 4 cts] No presentation on the 4th half-day
Which one would you like to set back ? (1-2 0-none) 1
In order to do that some constraints need to be removed:
PALM: Removing constraint Pm !== 4 from P4D
PALM: Removing constraint Mp !== 4 from P4D

!!! A solution has now been obtained
!!! (Am:1, Pm:2, Ma:3, Mp:4)
\end{verbatim}
}
\end{exemple}
\end{figure}

\subsubsection{Example}
Example~\ref{ex-use} shows \texttt{UFboxes} at use. As you can see
in that example, only understandable information is provided to
the user. In that example, Michael's representation of the
conference problem is used. When encountering a contradiction
(which shows that the problem is over-constrained), Michael is
confronted with an explanation of that contradiction. He chooses
to let Peter or Alan give a presentation before him (relaxing
block \texttt{PAB} in the example). Unfortunately, this will not
be sufficient\footnote{Remember that explanations cannot tell
exactly which constraint to remove but only focus on a set of
relevant constraints. } and Michael accepts to come on the fourth
half-day (relaxing block \texttt{N4D}). This time a solution is
obtained. Notice that only one constraint from this box needs to
be relaxed.

Example~\ref{ex-use} shows another feature of our problem. Once a
problem solved many user interactions have occurred and maybe
he/she wants to put back some relaxed constraints. \texttt{PaLM}
presents the set of relaxed \texttt{UFboxes} for reconsideration.
Here, Michael wants to put back the \texttt{PAB} block. A solution
is found. Notice that some further constraint relaxations are
needed (from the \texttt{N4D} box which is still relaxed).

\newpage
\section{Applications}
User-friendly explanations can be an invaluable tool in the
following situations:
\begin{itemize}
\item \textbf{Debugging} \\ Explanations can help focus on
relevant parts of the set of constraints when identifying a
contradiction. User-friendly ones are really necessary to interact
with the user: constraints sets need to get translated to all kind
of users.
\item \textbf{Solving over-constrained problems} \\ As we saw with our
toy example (the conference problem), user-friendly explanations
(used as in the debugging situation above) help the user
understanding the deep reasons of the lack of solution to his
problem. Moreover, user-friendly explanations are well suited for
distributed environments as in our example: a single explanation
is presented to different people who have different views on the
problem. The explanations is not modified, only the projection is
passed through the system. Notice that solving over-constrained
problems can be seen as debugging!
\item \textbf{Dynamic analysis of the solver's behavior} \\ As for
classic explanations, user-friendly explanations can explain
specific situations during search. Therefore, they can be used to
analyse (and report) the behavior of the solver to different kinds
of users: developer, end-users, managers, ...
\end{itemize}

\section{Related works}
\cite{goualard-sbox} introduced the notion of \emph{s-box} within
Constraint Logic Programming. \emph{s-boxes} are used to structure
the constraint store by considering sets of constraints as a
single one. It is worth noticing that \emph{s-boxes} have two main
drawbacks:
\begin{itemize}
\item considering a set of constraints as a single one is
relatively easy when considering numerical constraint: one just
needs to take the join of the projections. It is not that easy
with other kinds of constraints.
\item the main drawback relies in the behavior of \emph{s-boxes}.
Indeed, the solver behavior is modified since a whole set of
constraints is now replaced by a single one which means that
propagation stays into an \emph{s-box} until completion before
going to another one. This changes the solver behavior and
therefore \emph{s-boxes} may only be interesting for visualizing
the behavior of the solver or for identifying the reasons for a
contradiction but will be of no use when debugging a constraint
program.
\end{itemize}
Our proposal limits the grouping of constraints in an abstract
way. The concrete low-level constraints remain unmodified and
independent.

\cite{sqalli-inference} recently introduced user-friendly
explanations for logic puzzles. The idea here is to provide a
readable \emph{trace} of the solving mechanism by generating a
readable statement for each solver event. Generated explanations
are generally quite similar to hand-made ones although a bit
longer. However, explanations are associated to low-level
constraints and this work does not provide handling of sets of
constraints as a whole. Our proposal has that capability and
moreover can handle at the same time several views of a same
problem.

\section{Conclusion}
In this paper, we introduced the notion of user-friendly
explanations. The main idea is to consider constraint programs as
an hierarchy of constraints and to add information about that
hierarchy within the constraints. Therefore, users can be modelled
as a cut in the hierarchy tree and explanations can be projected
on their representation of the problems.

Our proposal has been implemented within the \texttt{PaLM} system
and shows interesting properties: possible handling of several
different users, adaptability to distributed systems, capability
of handling in a single box high-level constraints modelled as a
set of low-level constraints, complete generality of the approach,
...

Our current works include investigating real life use of our
user-friendly explanations. Our first experiment will be conducted
within the \texttt{ptidej} system \cite{albin-amiot-identifying}.

\section*{Acknowledgements}
This work originates from discussions and works with Olivier Lhomme.

\bibliographystyle{plain}
{\footnotesize

}
\end{document}